\documentclass[aps,prb,twocolumn,preprintnumbers,amsmath,amssymb]{revtex4}
\usepackage{graphicx}
\usepackage{xcolor}
\begin{document}
\title{Physical properties of (Mn$_{1-x}$Co$_{x}$)Si  at $x\simeq$0.06 – 0.1: quantum criticality matter.}

\author{A.~E.~Petrova}
\affiliation{P.~N.~Lebedev Physical Institute, Leninsky pr., 53, 119991 Moscow, Russia}
\affiliation{Institute for High Pressure Physics of RAS, Troitsk, Russia}
\author{S.~Yu.~Gavrilkin}
\affiliation{P.~N.~Lebedev Physical Institute, Leninsky pr., 53, 119991 Moscow, Russia}
\author{G.~V.~Rybalchenko}
\affiliation{P.~N.~Lebedev Physical Institute, Leninsky pr., 53, 119991 Moscow, Russia}
\author{Dirk Menzel}
\affiliation{Institut f\"{u}r Physik der Kondensierten Materie, Technische Universit\"{a}t Braunschweig, D-38106 Braunschweig, Germany}
\author{I.~P.~Zibrov}
\affiliation{Institute for High Pressure Physics of RAS, Troitsk, Russia}
\author{S.~M.~Stishov}
\email{stishovsm@lebedev.ru}
\affiliation{P. N. Lebedev Physical Institute, Leninsky pr., 53, 119991 Moscow, Russia}
\affiliation{Institute for High Pressure Physics of RAS, Troitsk, Moscow, Russia}

\begin{abstract}
We have grown and characterized three samples of Co doped MnSi and studied their physical properties (magnetization and magnetic susceptibility, heat capacity and electrical resistance). All three samples show non-Fermi liquid physical properties. From literature data  and current results follow that impurities (Co and Fe) eliminate the first order phase transition peaks and spread the fluctuation maxima in such a way that its low temperature part effectively reaches the zero temperature, where the fluctuations inevitably become quantum. The behavior of low temperature branches of the heat capacity of the samples suggests that a gradual transition from classical to quantum fluctuations can be described by a simple power function of temperature with the exponent less than one. The $d\rho/dT$ data generally support this suggestion. The values of the heat capacity exponents immediately lead to the diverging ratio $C_p/T$ and hence to the diverging effective electron mass. We found out that at large concentration of the dopant there are no distinct phase transition points. What we observe is a cloud of the helical fluctuations spreading over a significant range of concentrations and temperatures, which become quantum close to 0~K. 
\end{abstract}
\maketitle

\section{Introduction}
Recent years the quantum critical phenomena in magnetic systems have attracted much attention. In a course of intensive studies, the helical magnet MnSi has played a special role as a material available in an almost perfect single crystal form, whose phase diagram and physical properties at high pressure were well known. As was shown in many studies the magnetic phase transition point in MnSi decreases with pressure and reaches almost zero at about 1.4 GPa~\cite{1,2,3}. At the same time no evidence for quantum critical behavior were seen in MnSi at high pressure~\cite{4,5,6}. One of the reasons for this situation could be strong non hydrostatic stress developing at high compression~\cite{5,6,7}. To avoid this complication, it is reasonable to try using so called "chemical" pressure, which is replacing some atoms or ions of the material with chemically suitable atoms of smaller size. This procedure normally results in a volume decrease, which could imitate a high-pressure effect. Of course, it causes a certain disorder in materials that should be taken into account at the data interpretation. 
The described idea was used at studying the iron and cobalt doped of MnSi in Ref.~\cite{8,9,10}. As was shown in Ref.~\cite{9,10} the iron doped sample of MnSi with a nominal composition Mn$_{0.85}$Fe$_{0.15}$Si reveals features inherited to the proximity of quantum critical point. However, the results for Co doped samples, obtained in Ref.~\cite{9}, were not quite conclusive.

So, with all that in mind we decided to extend the study~\cite{9} in relation of Co doped samples of MnSi. 

\section{Experimental}
We prepared three samples with the nominal composition:  Mn$_{0.94}$Co$_{0.06}$Si, Mn$_{0.92}$Co$_{0.08}$Si, Mn$_{0.9}$Co$_{0.1}$Si.  Polycrystalline ingots containing Mn (Chempur; purity 99.99\%), Co (Chempur; purity 99.95\%), and Si ($\rho_n$=300 Ohm cm, $\rho_p$=3000 Ohm cm) were prepared by arc melting under argon atmosphere. Afterwards, single crystals have been grown using the triarc Czochralski technique.

The electron-probe microanalysis shows that real composition are: Mn$_{0.93}$Co$_{0.057}$Si, Mn$_{0.92}$Co$_{0.063}$Si, Mn$_{0.89}$Co$_{0.09}$Si, which indicates that two first samples have practically identical compositions despite the initial concentration of doping element (Table~\ref{tab:table1}).

\begin{table}
\caption{\label{tab:table1} Chemical compositions and lattice parameters of (Mn,Co)Si, (Mn,Fe)Si and MnSi samples. $a$-lattice parameter, $V$-unit cell volume.}
\begin{ruledtabular}
 \begin{tabular}{cccc}
Nominal& Electron-probe& $a$& $V$ \\
composition &Composition & \AA & \AA$^3$ \\ 
\hline 
Mn$_{0.94}$Co$_{0.06}$& Mn$_{0.93}$Co$_{0.057}$Si\footnotemark[1] & 4.5522 & 94.333 \\ 
\hline 
Mn$_{0.92}$Co$_{0.08}$Si & Mn$_{0.92}$Co$_{0.063}$Si\footnotemark[1] & 4.5519 & 94.313 \\ 
\hline 
Mn$_{0.9}$Co$_{0.1}$Si & Mn$_{0.89}$Co$_{0.09}$Si\footnotemark[1] & 4.5499 & 94.189 \\ 
\hline 
Mn$_{0.85}$Fe$_{0.15}$Si & Mn$_{0.83}$Fe$_{0.17}$Si\footnotemark[2] & 4.5462 & 93.961 \\ 
\hline 
MnSi & MnSi\footnotemark[3] & 4.5598 & 94.8063 \\ 
\end{tabular}
\end{ruledtabular}
\footnotetext[1]{Current paper.}
\footnotetext[2]{Ref.~\cite{10},  (Chemical composition data in the table obtained in new facilities are slightly different from Ref.~\cite{10})  }
\footnotetext[3]{Ref.~\cite{11}.}
\end{table}

We performed some magnetic, heat capacity and resistivity measurements to characterize the (Mn,Co)Si samples. All measurements were made making use the  Quantum Design PPMS system with the heat capacity and vibrating magnetometer moduli and the He-3 Refrigerator. The resistivity data were obtained with the standard four terminals scheme using the spark welded Pt wires as electrical contacts.

\begin{figure}[htb]
\includegraphics[width=80mm]{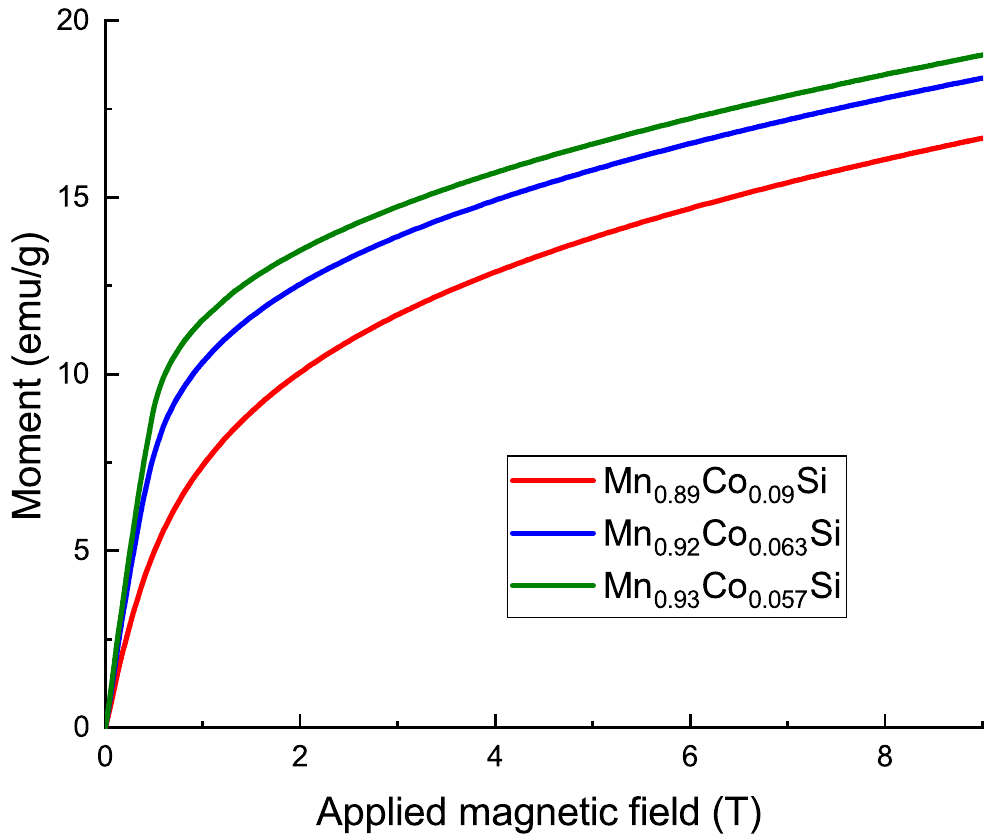}
\caption{\label{fig1} (Color online) Magnetizations of samples (Mn,Co)Si at 2 K.}
\end{figure}

\begin{figure}[htb]
\includegraphics[width=80mm]{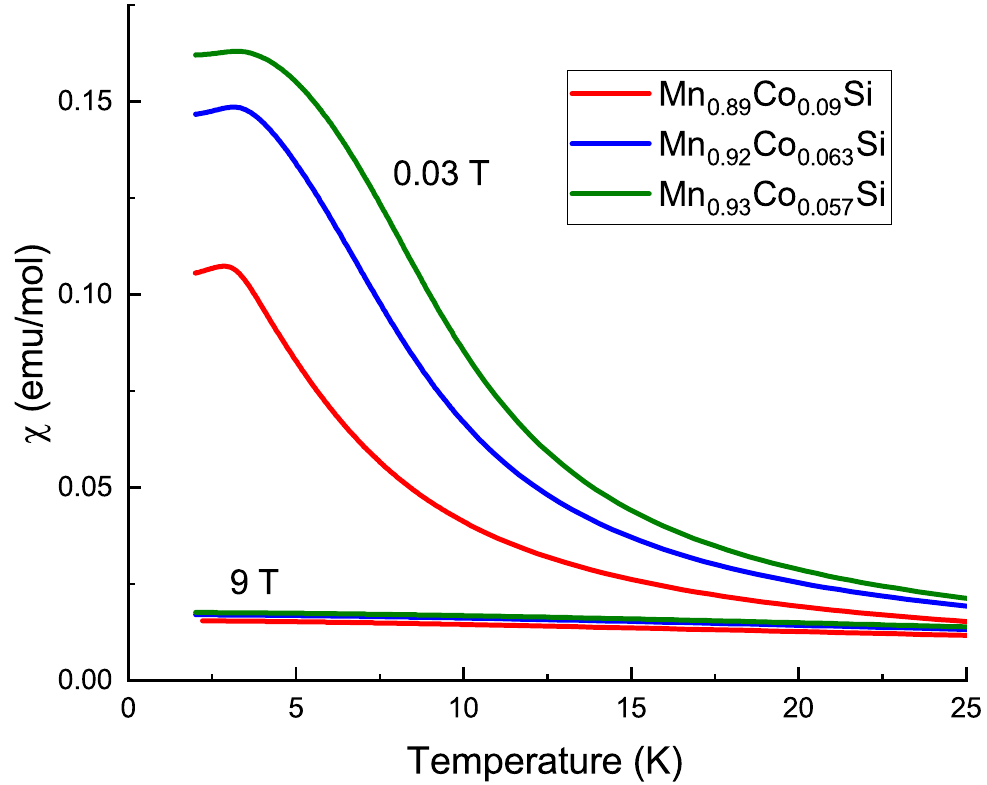}
\caption{\label{fig2} (Color online) Magnetic susceptibility of (Mn,Co)Si as a function of temperature at 0.03 and 9 T.}
 \end{figure}

\begin{figure}[htb]
\includegraphics[width=80mm]{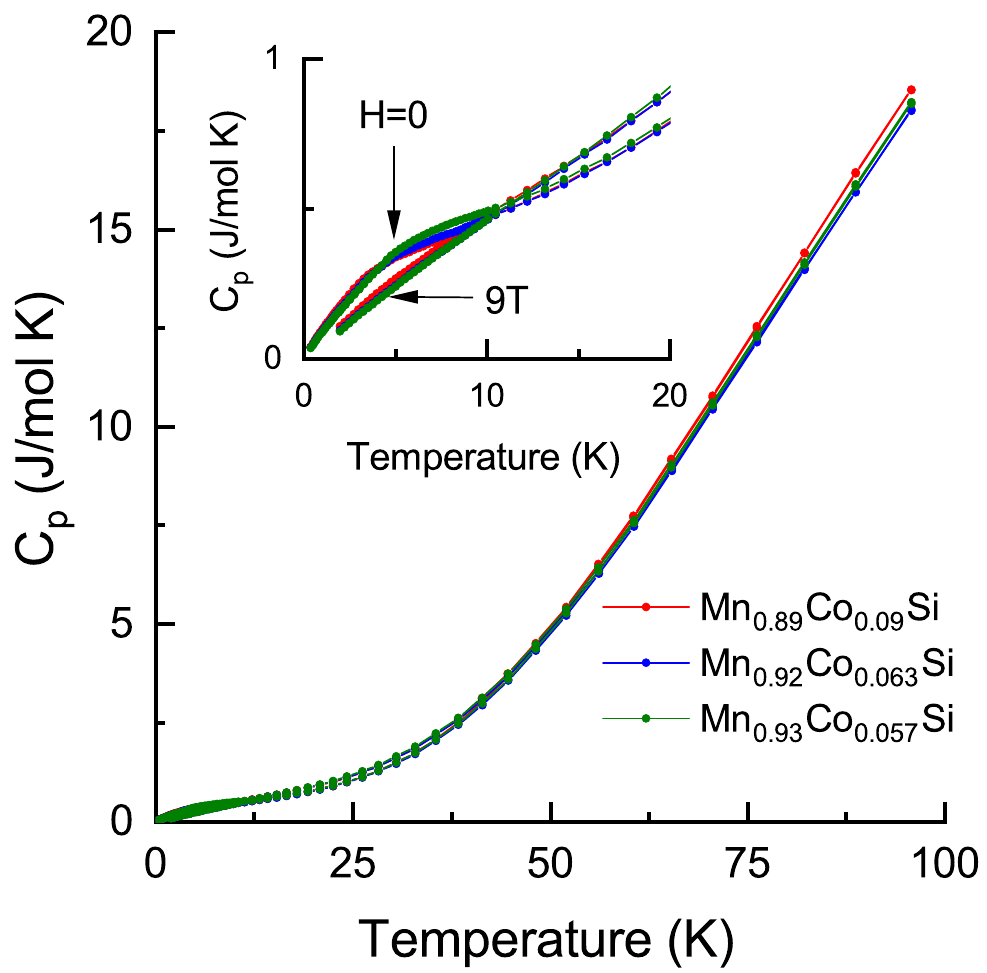}
\caption{\label{fig3} (Color online) Heat capacity of (Mn,Co)Si at zero and 9 T as a function of temperature in the range 0.4--100 K. Enlarge part of the plot is in the inset. }
\end{figure}

\begin{figure}[htb]
\includegraphics[width=80mm]{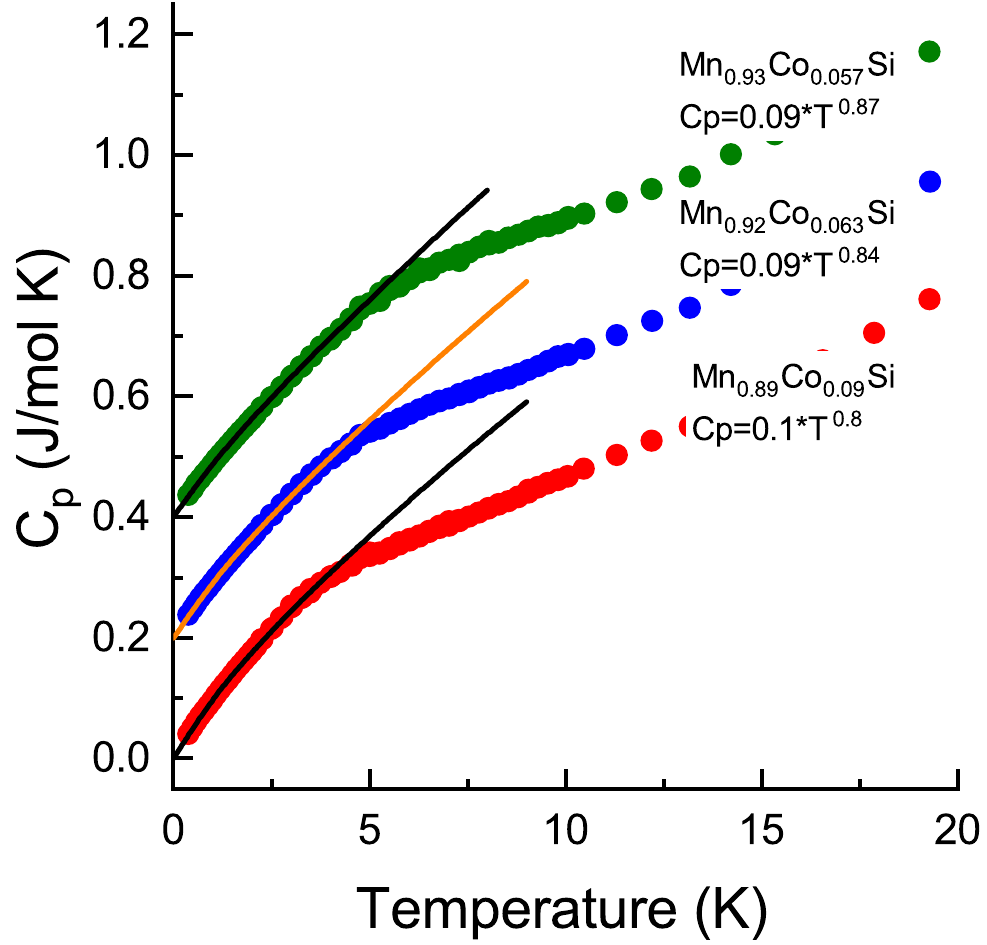}
\caption{\label{fig4} (Color online) Illustration to the fitting of the low temperature part of heat capacity to the power function. The values of the power exponents shown in the plot. The data are shown with offsets for better viewing.}
\end{figure}

\begin{figure}[htb]
\includegraphics[width=80mm]{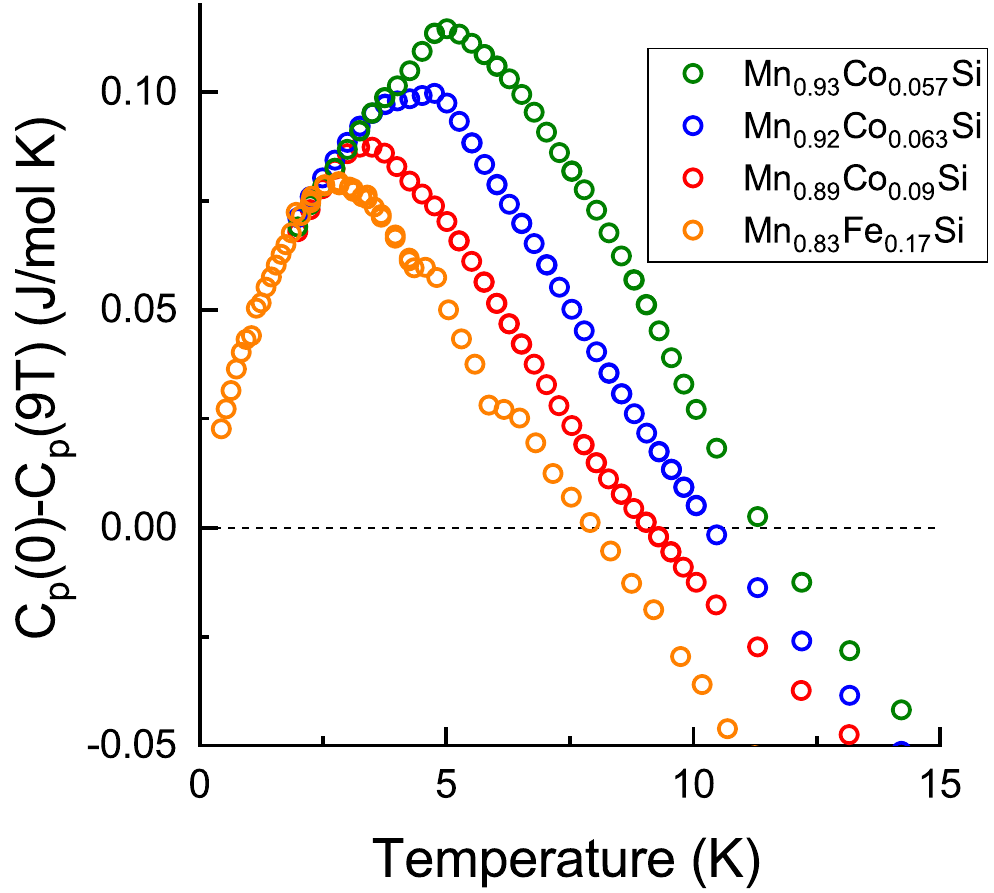}
\caption{\label{fig5} (Color online) The difference between heat capacity at zero magnetic field $C_p(0)$ and heat capacity at 9~T ($C_p(9 T)$ for (Mn,Co)Si samples (current work) and one sample of (Mn,Fe)Si~\cite{10}.}
\end{figure}

\begin{figure}[htb]
\includegraphics[width=80mm]{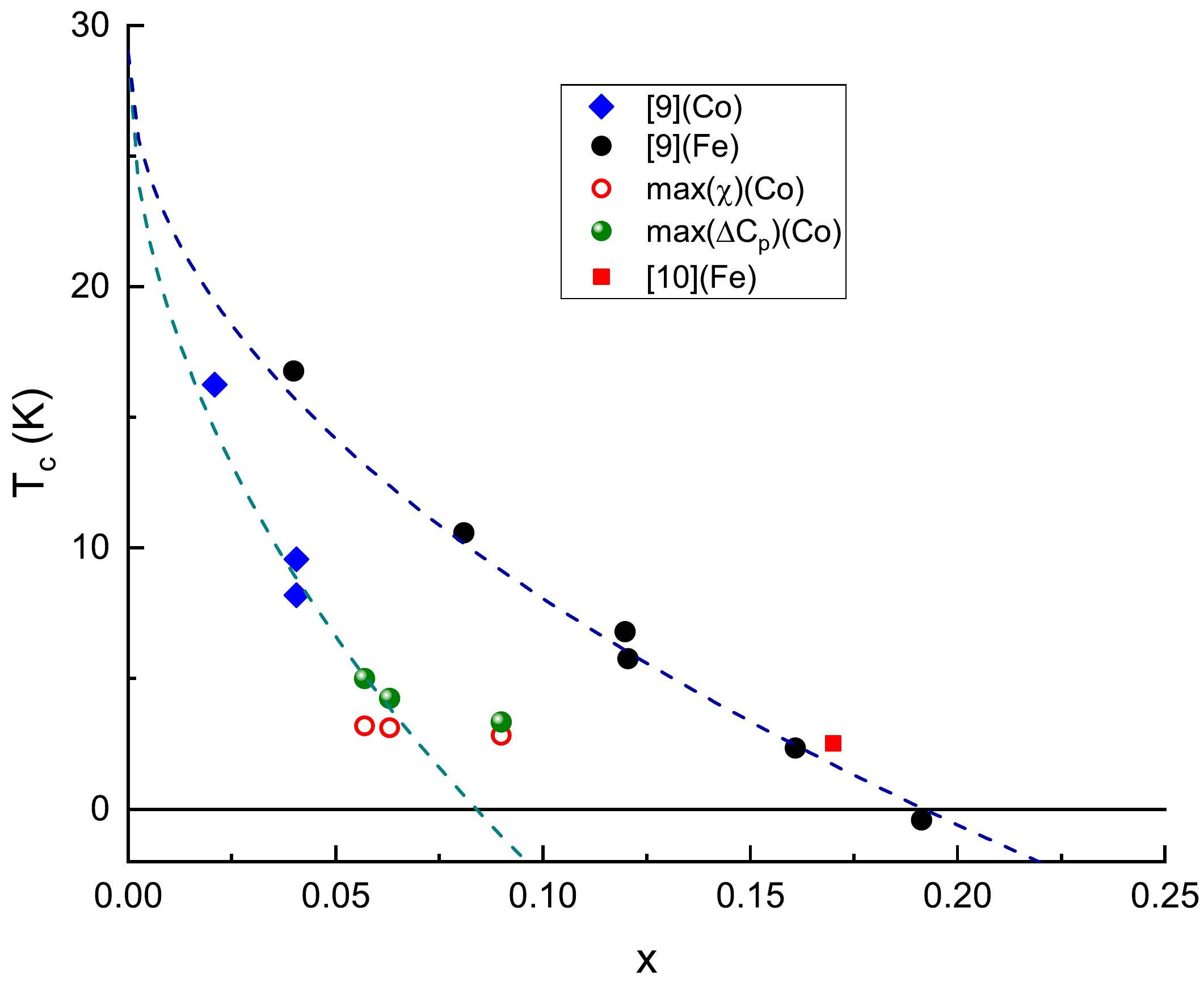}
\caption{\label{fig6} (Color online) "Phase transition" temperature as a function of concentrations for Mn$_{1-x}$Fe$_x$Si~\cite{9,10} and Mn$_{1-x}$Co$_x$Si~\cite{9} and current data. Indications of chemical elements in legend  show nature of dopants.   In Ref.~\cite{9} "phase transition" temperatures were determined from the temperature dependence of magnetic moments. Current data were taken as the $\chi$ maxima (Fig.~\ref{fig2}) and maxima of $\Delta C_p (0-9T)$ (Fig.~\ref{fig5}). Note that the square datapoint on the "iron" curve corresponds to the sample studied in~\cite{10} (Corrected composition was used. Transition temperature was taken as a maximum of $\Delta C_p (0-9T)$(Fig.~\ref{fig5})).
} 
\end{figure}

\begin{figure}[htb]
\includegraphics[width=80mm]{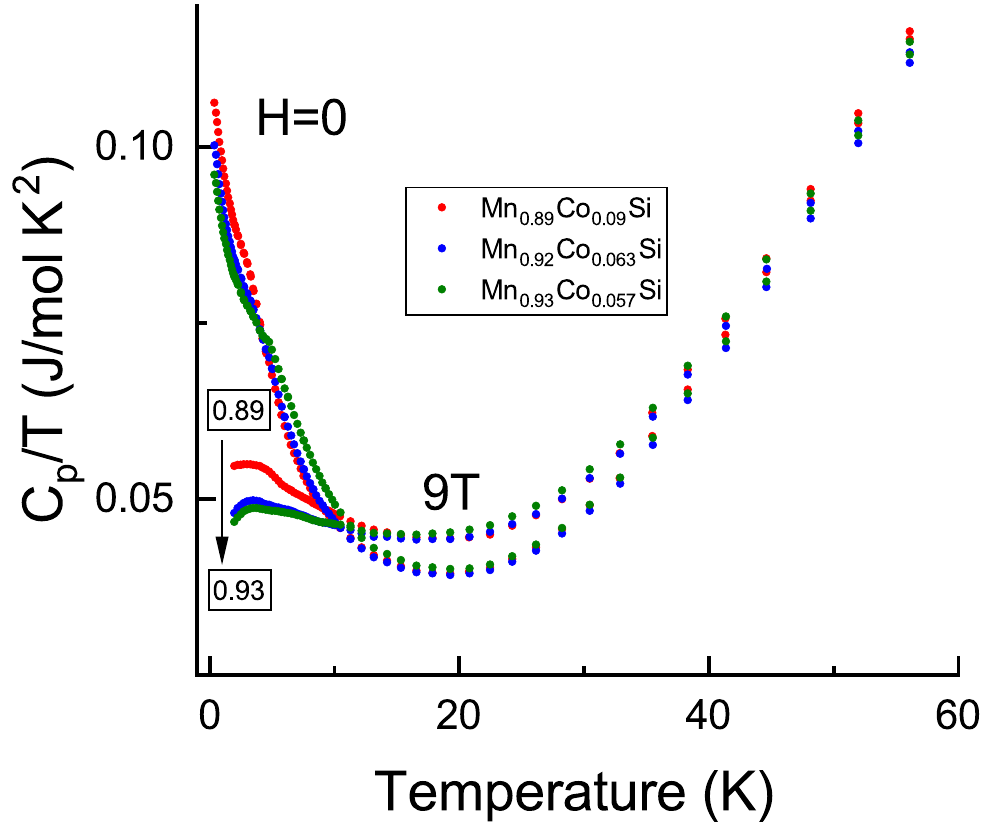}
\caption{\label{fig7} (Color online) The ratio $C_p/T$ for (Mn,Co) samples as a function of temperature at zero and 9~T magnetic fields. Is seen that diverging of $C_p/T$ is suppressed by strong magnetic field. } 
\end{figure}

\begin{figure}[htb]
\includegraphics[width=80mm]{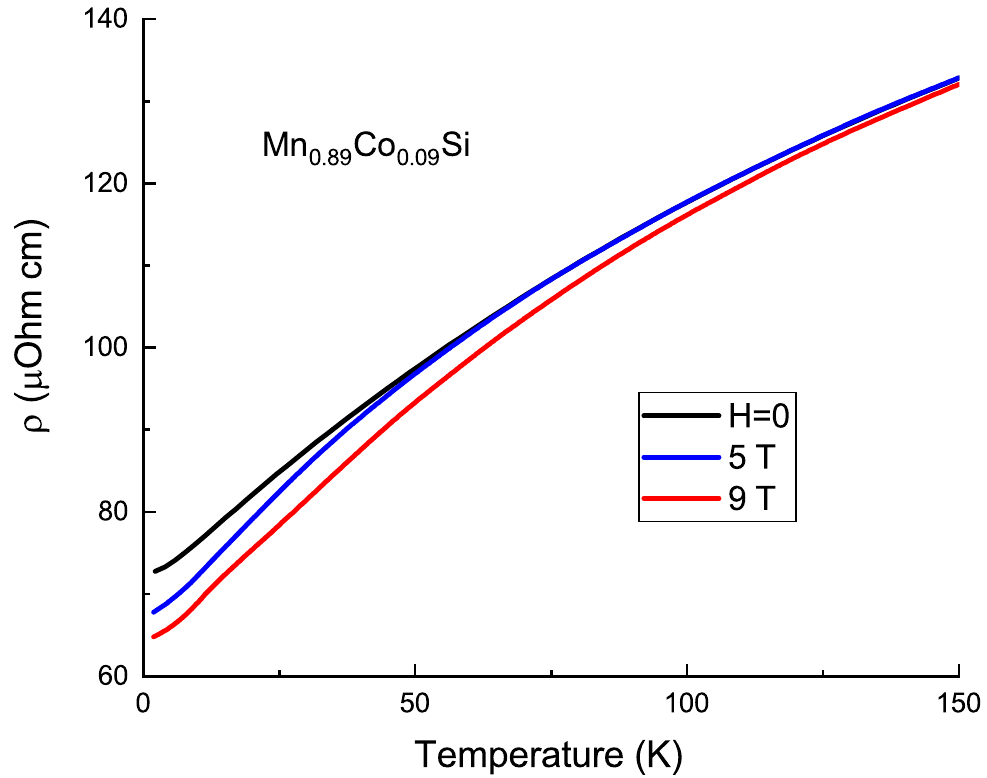}
\caption{\label{fig8} (Color online) The dependence of resistivity of Mn$_{0.89}$Co$_{0.09}$Si on temperature at various magnetic fields.} 
\end{figure}

The experimental data are shown in Fig.~\ref{fig1}--\ref{fig8}.

Fig.~\ref{fig1} illustrates magnetization of the samples in the fields to 9 T. As is seen there are no indications for spontaneous magnetic moments in these materials.

The magnetic susceptibility of (Mn,Co)Si samples is demonstrated in Fig.~\ref{fig2}. in the limited range of temperature for a better viewing of the specific features of these cobra-like curves. At the first sight these cobra-like features are nothing else than strongly deformed maximum of $\chi (T)$ observed in the pure MnSi at the phase transition point. Similar maxima in $\chi (T)$ were seen in~\cite{9} at lower Co concentrations. If these maxima indeed correspond to the "phase transition" points it worth comparing the data with the results of Ref.~\cite{9} that will be done lately in this paper.

The results of heat capacity measurements are displayed in Fig.~\ref{fig3}. As is seen in Fig.~\ref{fig4} the lines of $C_p (T)$ changes their slope rather sharply at temperatures in the vicinity of 4-5~K. Actually, these features can be attributed to the strongly smeared out the fluctuation induced heat capacity maximum associated with the magnetic phase transition in pure MnSi.

The mentioned slope change, which becomes more evident after a subtraction from the heat capacity curve at zero magnetic field the corresponding curve at 9 T, occurs at $\sim$5~K (Mn$_{0.93}$Co$_{0.057}$Si), $\sim$4.7~K (Mn$_{0.92}$Co$_{0.063}$Si) and $\sim$3.4~K (Mn$_{0.89}$Co$_{0.09}$Si) (Fig.~\ref{fig5}). Actually, this manipulation implies a substruction of some of background contributions, including phonon one to the heat capacity leaving its fluctuation part intact. At the same time as seen in Fig.~\ref{fig5} this procedure reduces the low temperature branch of the $C_p$ to the puzzling universal line even including data on MnFeSi~\cite{10}. On the other hand, the high temperature branches of the differential $C_p$ curves became negative at some temperatures therefore supporting our old conclusion that strong fluctuations give negative contributions to the heat capacity of paramagnetic phase of MnSi~\cite{11}.
 
\begin{figure}[htb]
\includegraphics[width=80mm]{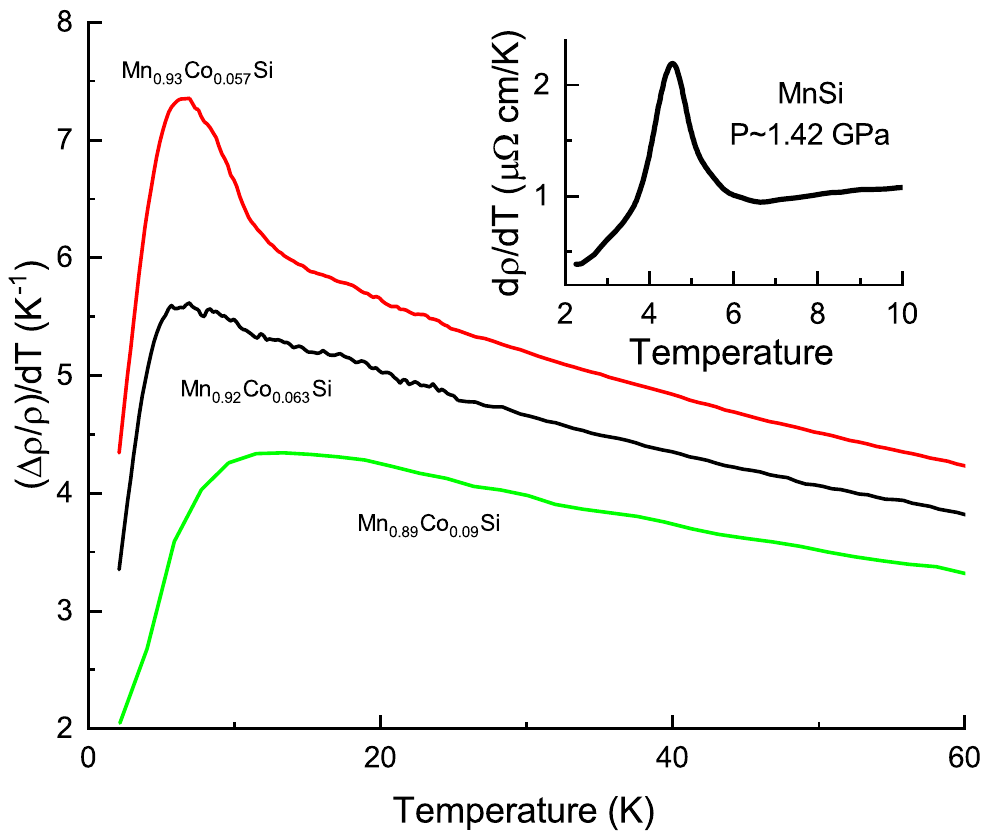}
\caption{\label{fig9} (Color online) Temperature derivatives of reistivity of the samples Mn$_{1-x}$Co$_x$Si. In the inset $d\rho/dT$ data for pure MnSi at the phase transition occurring at high pressure are shown (after Ref.~\cite{15}). } 
\end{figure}

\begin{figure}[htb]
\includegraphics[width=80mm]{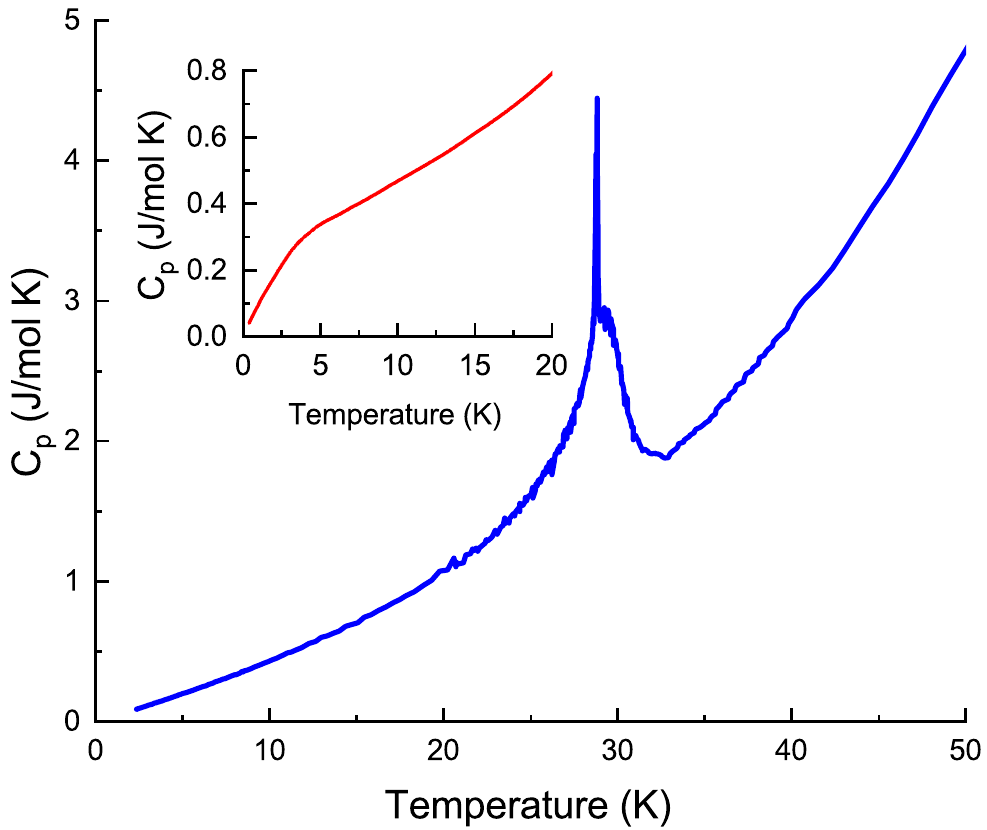}
\caption{\label{fig10} (Color online) Temperature dependence of heat capacity near the magnetic phase transition in MnSi (after Ref.~\cite{15}). In the inset heat capacity anomaly in Mn$_{0.89}$Co$_{0.09}$Si is shown. } 
\end{figure}

\begin{figure}[htb]
\includegraphics[width=80mm]{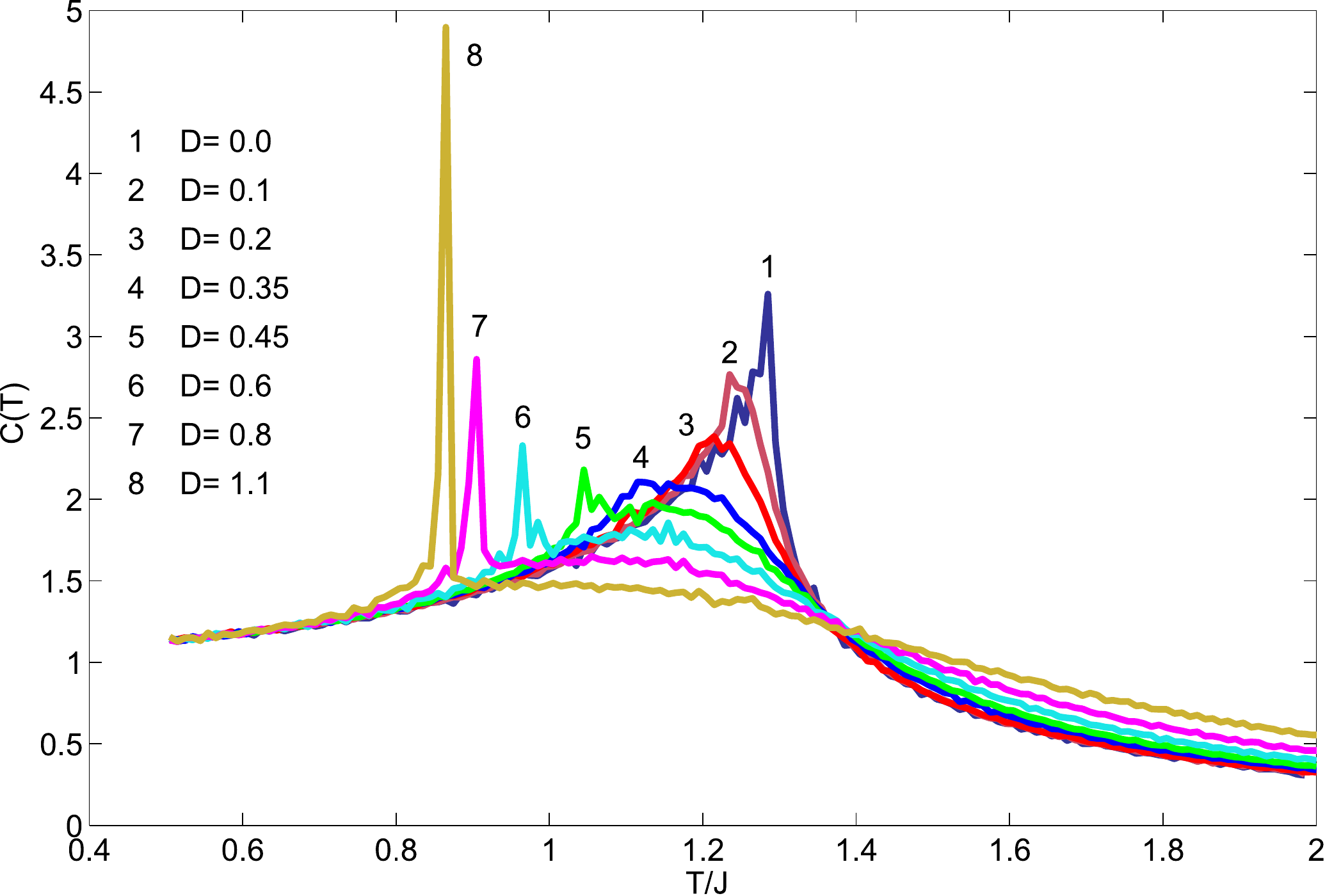}
\caption{\label{fig11} (Color online) Temperature dependence of the specific heat $C(T)$ for the Hamiltonian $H=H_J+H_D$ for different values of the DM (D)interaction (after Ref.~\cite{16}).} 
\end{figure}

\begin{figure}[htb]
\includegraphics[width=80mm]{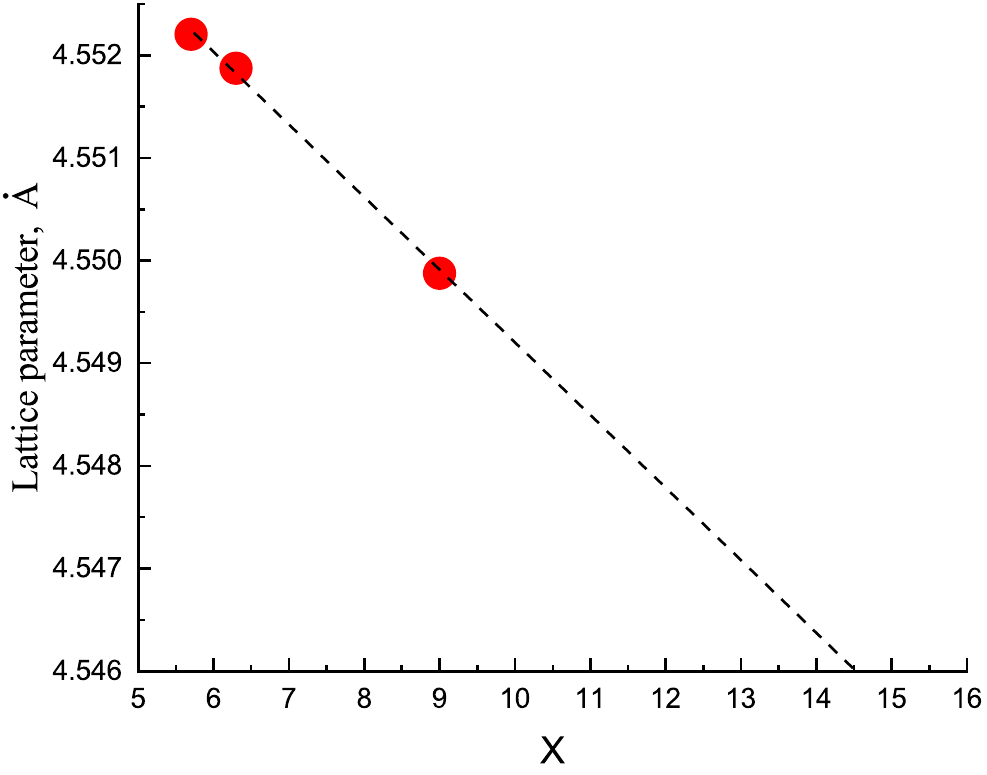}
\caption{\label{fig12} (Color online) Lattice parameter of Mn$_{1-x}$Co$_x$Si as a function of Co concentration (X) (Table~\ref{tab:table1}).} 
\end{figure}

The maximum temperatures of the curves in Fig.~\ref{fig5} generally correlate with the local maxima of $\chi(T)$ taken as the "phase transition" temperatures (see Fig.~\ref{fig2}). In Fig.~\ref{fig6} is seen that data points with concentration $x\approx 0.06$ fit well with data of Ref.~\cite{9}, whereas the ones with $x=0.09$ sharply deviates from the suggested curve. Moreover, an impression arises that the "phase transition" temperature somehow avoids declining at further increasing of Co concentration.

Anyway, the low temperature branches of the heat capacity curves reflect a gradual transition from classical to quantum fluctuations, which can be described by a simple power function of temperature with the exponent less than one. The latter immediately leads to the diverging ratio $C_p/T$ (see Fig.~\ref{fig7}) and hence to the diverging effective electron mass, which is a signature of the quantum critical behavior.

The resistivity of the samples as a function of temperature exhibits an evolution from low temperature region of not quite clear underlying physics to the "saturation" regime typical of the strongly disordered metals~\cite{12} (Fig.~\ref{fig8} (see also Ref.~\cite{10}).

Fig.~\ref{fig8} illustrates this kind of the resistivity behavior in magnetic field to 9~T using the sample Mn$_{0.89}$Co$_{0.09}$Si as a typical example. In attempt to understand the low temperature conduct of the resistivity we performed a fitting of the corresponding data with the expression $\rho=A+BT^n$ in the range 2-8~K. The following values of exponents $n$ were obtained: $n=1.7~(H=0), n=1.4~(5~T), n=1.6~ (9~T)$, which uncovers non-Fermi-liquid behavior. This finding adds one more puzzle to the problem of non-Fermi-liquid exponent~\cite{12,13}.

Temperature derivatives of reistivity of the samples Mn$_{1-x}$Co$_x$Si. Are displayed in Fig.~\ref{fig9}.
These data correlate in certain aspects with those shown in Fig.~\ref{fig4},  \ref{fig5} identifying the low temperature parts of the corresponding curves with the smeared out low temperature sides of the fluctuation maxima (see the inset where $d\rho/dT$ data for the high pressure phase transition in pure MnSi are demonstrated).

\section{Discussion}
First, let us to turn attention to Fig.~\ref{fig10}, demonstrating a behavior of heat capacity in pure MnSi at the magnetic phase transition~\cite{15}. As can be seen in Fig.~\ref{fig10} the phase transition in MnSi is characterized by a sharp peak on top of a rounded maximum forming a shoulder in the high temperature side of the peak. According to the Monte Carlo calculations ~\cite{16} such structure of the heat capacity results from the perturbation of a virtual second order ferromagnetic phase transition by the helical fluctuations, which smeared out the transition and eventually condense into the helically ordered phase (see Fig.~\ref{fig11}).

From the data~\cite{9} and the current results follow that impurities (Co and Fe) extinguish the first order peaks and spread the fluctuation maxima in such a way that its low temperature part effectively reaches the zero temperature, where the fluctuations inevitably become quantum (see Fig.~\ref{fig5}). Moreover, a doping action of Co, obviously saturates at some concentrations, so the $\chi$ and $C_p(0)-C_p(9T)$ maxima do not move significantly with concentration increasing (Fig.~\ref{fig2},~\ref{fig5}). Note that in Ref.~\cite{10} the quantum critical trajectory was identified for the sample Mn$_{0.83}$Fe$_{0.17}$Si with a lattice parameter 4.5462~\AA, the value, which coincides with the lattice parameter of pure MnSi at the phase transition point at $T\sim 0$~K and $P\sim 14.5$~kbar.  Follow this reason, we plot available lattice parameter data of (Mn,Co)Si as a function of Co concentration (Fig.~\ref{fig12}). The extrapolation by the Vegard's rule shows that the value of lattice parameter equal to 4.5462~\AA \  for (Mn,Co)Si can be attained at 14-15 \% concentration of Co i.e. at composition Mn$_{0.85}$Co$_{0.14-0.15}$Si. However, it is not clear whether this concentration is reachable. Moreover, in the light of the current study one hardy can find a unique single critical trajectory. Probably the same should be applicable to the Mn$_{1-x}$Fe$_x$Si case. So, the conclusion of Ref.~\cite{10} may be need a modification.

Thus, at large concentrations of the dopant there is no a definite phase transition point, instead one has a cloud of the helical fluctuations, which spread over a significant range of concentrations and temperatures and become quantum close to 0 K. Diverging the ratio $C_p/T$ justifies this fact and leads to infinite electronic effective mass as a signature of quantum criticality (Fig.~\ref{fig7}).

\section{Conclusion}
   
We have grown and characterized three samples of Co doped MnSi and studied their physical properties (magnetization and magnetic susceptibility, heat capacity and electrical resistance). All three samples show non-Fermi liquid physical properties. From  the data~\cite{9} and current results follow that impurities (Co and Fe) eliminate the first order phase transition peaks and spread the fluctuation maxima in such a way that its low temperature part effectively reaches the zero temperature, where the fluctuations inevitably become quantum (see Fig.~\ref{fig4},\ref{fig5}). The behavior of low temperature branches of the heat capacity of the samples (Fig.~\ref{fig4},\ref{fig5}) suggests that a gradual transition from classical to quantum fluctuations can be described by a simple power function of temperature with the exponent less than one. The $d\rho/dT$ data  generally support this suggestion (Fig.~\ref{fig12}). The value of the heat capacity exponents immediately leads to the diverging ratio $C_p/T$ (see Fig.~\ref{fig7}) and hence to the diverging effective electron mass. We found out that at large concentration of the dopant there are no distinct phase transition points (Fig.~\ref{fig2},\ref{fig5}). What we observe is a cloud of the helical fluctuations spreading over a significant range of concentrations and temperatures, which become quantum close to 0 K (Fig.~\ref{fig6}). 

\section{Acknowledgements}
The authors gratefully acknowledge the technical support of Dr. V.A. Sidorov. 
AEP and SMS greatly appreciate financial support of the Russian Foundation for Basic Research (grant No. 18-02-00183) and the Russian Science Foundation (grant 17-12-01050).

\end{document}